\documentclass{Interspeech}

\interspeechcameraready

\title{LLM-based phoneme-to-grapheme for phoneme-based speech recognition
\thanks{*Corresponding author. This work is partly supported by Guangxi Science and Technology Project (2022AC16002) and National Natural Science Foundation of China (62466055). The code, models and data for LLM-P2G are released at https://github.com/thu-spmi/CAT/blob/master/docs/whatsnew.md}
}

\author[affiliation={1}]{Te}{Ma}
\author[affiliation={2}]{Min}{Bi}
\author[affiliation={1}]{Saierdaer}{Yusuyin}
\author[affiliation={1}]{Hao}{Huang}
\author[affiliation={3}]{Zhijian}{Ou$^{*}$}

\affiliation{School of Computer Science and Technology}{Xinjiang University}{China}
\affiliation{Guangxi Radio and Television Monitoring Center}{Guangxi}{China}
\affiliation{Speech Processing and Machine Intelligence (SPMI) Lab}{Tsinghua University}{China}

\email{huanghao@xju.edu.cn, ozj@tsinghua.edu.cn}


\usepackage{comment, multirow}

\begin{document}

\maketitle

\begin{abstract}
    
In automatic speech recognition (ASR), phoneme-based multilingual pre-training and crosslingual fine-tuning is attractive for its high data efficiency and competitive results compared to subword-based models. However, Weighted Finite State Transducer (WFST) based decoding is limited by its complex pipeline and inability to leverage large language models (LLMs). Therefore, we propose LLM-based phoneme-to-grapheme (LLM-P2G) decoding for phoneme-based ASR, consisting of speech-to-phoneme (S2P) and phoneme-to-grapheme (P2G). 
A challenge is that there seems to have information loss in cascading S2P and P2G.
To address this challenge, we propose two training strategies: data augmentation with noisy phonemes (DANP), and randomized top-$K$ marginalized (TKM) training and decoding.
Our experimental results show that LLM-P2G outperforms WFST-based systems in crosslingual ASR for Polish and German, by relative WER reductions of 3.6\% and 6.9\% respectively.
\end{abstract}

\section{Introduction}

\begin{figure}[t]
\centering
\includegraphics[width=\linewidth]{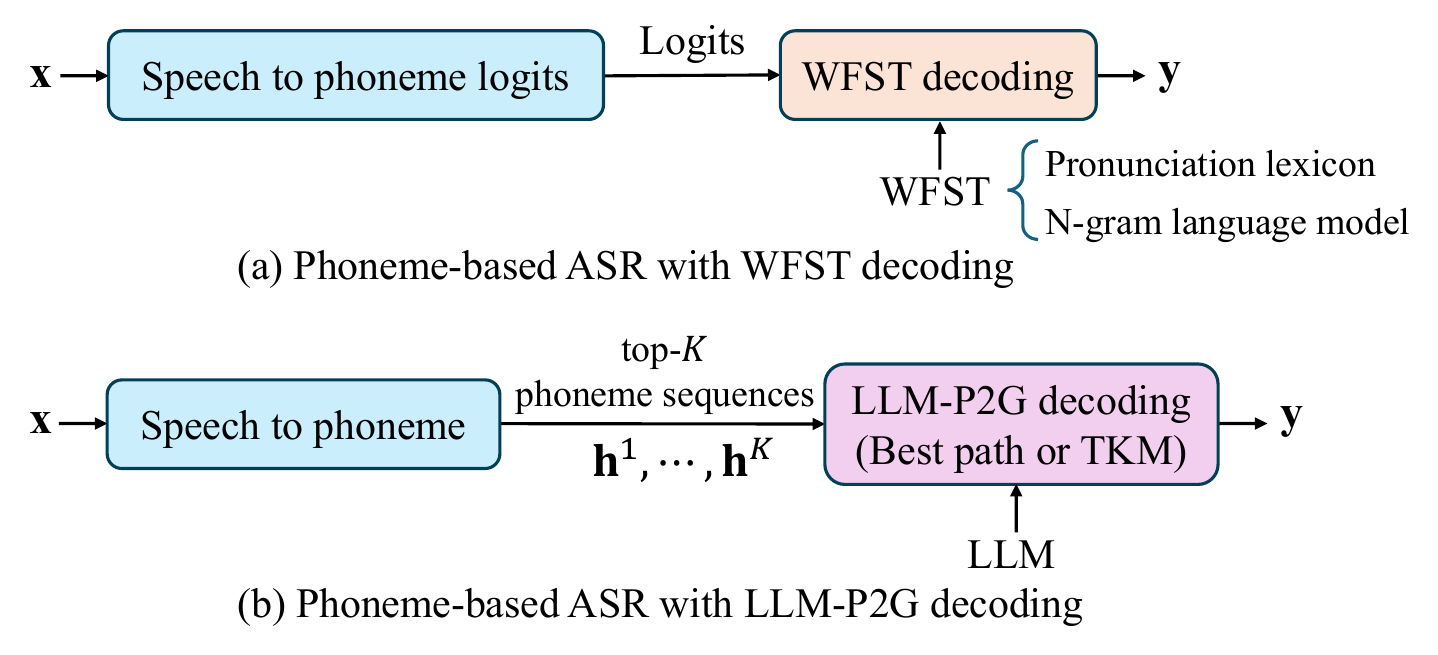} 
\caption{Phoneme-based ASR with WFST-based decoding (a) and with LLM-P2G decoding (b).
In recognizing speech $x$ into text $y$, phonemes arise as intermediate states, denoted by $h$.
LLM-P2G decoding can be either best path decoding or top-$K$ marginalized (TKM) decoding.
}
\label{fig:overview}
\vspace{-2.2em}
\end{figure}

Most languages worldwide are under-resourced, posing significant challenges in developing high-performance ASR systems. Therefore, multilingual pre-training and crosslingual fine-tuning have been developed, enabling information sharing and knowledge transferring between languages \cite{li2020universal, zhu2021multilingual, tjandra2023massively, xu2021simple, yusuyin2023investigation, glocker2023allophant}. Among these advancements, phoneme-based multilingual pre-training and crosslingual fine-tuning, in particular the weakly-phonetic-supervision-based approach, called Whistle \cite{yusuyin2025whistle,dong2024whistle}, is attractive for its high data efficiency and competitive results compared to subword-based models. 
Presumably, this is because phoneme-based supervision enables more efficient data sharing than subword-based supervision.
Phonetic units such as described in International Phonetic Alphabet (IPA) are exactly those sounds shared in human languages.
However, challenges remain for decoding in phoneme-based ASR. The widely used Weighted Finite State Transducer (WFST) \cite{mohri2008speech} based decoding (Figure \ref{fig:overview}(a)) delivers strong decoding performance but has two major drawbacks: 1) it involves a complex pipeline, which needs construction of pronunciation lexicons and compiling of WFSTs; 2) it is not easy to effectively leverage the rich linguistic knowledge in large language models (LLMs) \cite{xue-etal-2021-mt5,achiam2023gpt,touvron2023llama}.

In this work, we propose LLM-based phoneme-to-grapheme (referred to as \emph{LLM-P2G}) for phoneme-based ASR, as shown in Figure \ref{fig:overview}(b).
It belongs to a two-step ASR architecture \cite{xue2023tranusr,twostage}, which can be referred to as SPG, consisting of speech-to-phoneme (S2P) and phoneme-to-grapheme (P2G).
The S2P model can be obtained by fine-tuning a phoneme-based multilingual S2P backbone (Whistle) over speech data with phoneme labels \cite{yusuyin2025whistle}, which we refer to as \emph{Whistle-S2P}.
After converting speech into phonemes by the S2P model, the decoding of phonemes into text (usually represented by subwords) is called phoneme-to-grapheme.
The P2G model can be obtained by fine-tuning an LLM. Since both phonemes and subwords are discrete tokens\footnote{Remarkably, the IPA symbols all fall in the token set of mT5 -- the LLM used in our experiments.}, such P2G model can be naturally trained over LLMs, and inherits the powerful language understanding and generation capabilities of LLMs.
Interestingly, it is found that in large-scale models, the P2G capability (called IPA transliterate in \cite{wei2022emergent}) emerges.

A challenge in building the two-step ASR is that there seems to have information loss in cascading S2P and P2G.
In decoding, the hypothesized phoneme sequence output from S2P may contain errors, which may propagate to P2G.
To address this challenge, we propose two training strategies: data augmentation with noisy phonemes (DANP), and randomized top-$K$ marginalized (TKM) training and decoding.
In DANP, we run S2P to generate more diversified hypothesized phoneme sequences (i.e. adding noise), which are used to train the P2G model.
This makes the training and decoding of P2G to be more matched.
In TKM, marginalization over hypothesized phoneme sequences is performed, which reduces the influence of decoding with only a single hypothesized phoneme sequence.
These methods effectively overcome the possible information loss and successfully optimize the P2G model.
Our experiments show that using the same S2P model, LLM-P2G decoding surpasses WFST-based decoding in crosslingual ASR tasks for Polish and German, achieving up to 3.6\% and 6.9\% relative WER reductions, respectively. 


\section{Related Work}

The two-step idea of recognizing speech to phonemes and then to graphemes has been studied for crosslingual ASR \cite{xue2023tranusr,twostage}. These prior works share a similar motivation with ours that phoneme-based supervision is advantageous for multilingual acoustic representation learning. However, both studies do not explore using LLMs for P2G for phoneme-based ASR.

Large language models (LLMs) \cite{xue-etal-2021-mt5,achiam2023gpt,touvron2023llama} have demonstrated remarkable performance on a variety of natural language processing (NLP) tasks, showcasing strong capabilities in understanding and generating human languages. 
Different interfaces between speech and languages have been studied to integrate LLMs with ASR, including ASR-generated text and continuous embeddings of speech. 
In \cite{ma2023can}, ASR-generated text is fed into the LLM for error correction.
In \cite{wu2023decoder, chu2023qwen, ma2024embarrassingly}, the output from speech encoders are connected to LLMs through projector modules.
However, these prior studies are generally tested for monolingual ASR (mainly English) and not aimed to improve phoneme-based speech recognition for low-resourced languages.
The phoneme-based ASR architecture with Whistle-S2P and LLM-P2G is investigated in this work and is found to be efficient in leveraging large acoustic and language pre-trained models for speech recognition.
In contrast, using ASR-generated text as the interface is heavier than using phonemes, and using continuous embeddings of speech as the interface needs additional projection modules.



\section{Method}

\subsection{ASR architecture}

In the following, we introduce different ASR architectures, including the traditional one and the new architecture with LLM-P2G proposed in this work.
For both architectures, the acoustic model (or say S2P) can be obtained by fine-tuning a phoneme-based multilingual S2P backbone (Whistle) over speech data with phoneme labels \cite{yusuyin2025whistle}, which we refer to as \emph{Whistle-S2P}.

\textbf{Phoneme-based or subword-based ASR with WFST decoding} is a traditional architecture, as shown in Figure 1(a). The WFST framework is widely used in phoneme-based or subword-based ASR systems, which integrates the constraints of the acoustic model, the lexicon, and the language model into a unified graph structure. 
An acoustic encoder is used to convert speech features into phoneme or subword logits. Then, WFST based decoding is used to decode logits into words. The WFST framework can also be adopted for subword-based ASR, where subwords are used as modeling units instead of phonemes, and the lexicon maps word to subwords.

\textbf{Phoneme-based ASR with LLM-P2G decoding} is proposed in this work, as shown in Figure 1(b). 
Given acoustic observation $\mathbf{x} \triangleq x_1,\cdots,x_T$, the task of ASR is to find the most likely text $\mathbf{y} \triangleq y_1,\cdots\,y_L$, represented by subword sequence.
In phoneme-based ASR, the hypothesized phoneme sequence is denoted by $\mathbf{h}$, and a two-step architecture can be defined as follows, where phonemes arise as latent variables:
\begin{equation} 
\label{eq:S2P2G}
\begin{split}
p(\mathbf{y}|\mathbf{x}) = \sum_{\mathbf{h}} p(\mathbf{h}|\mathbf{x})p(\mathbf{y}|\mathbf{h})
\end{split}
\end{equation}
Here we assume that when given phonemes, we can infer text from phonemes, without depending on lower-level speech signal, i.e. $p(\mathbf{y}|\mathbf{x}, \mathbf{h})=p(\mathbf{y}|\mathbf{h})$.
Thus, we obtain a two-step ASR architecture, which can be referred to as SPG, consisting of a S2P model $p(\mathbf{h}|\mathbf{x})$ and a P2G model $p(\mathbf{y}|\mathbf{h})$. 

The P2G model can be obtained by fine-tuning an LLM, which we refer to as \emph{LLM-P2G}.
An LLM is a sequence generation model trained on massive amounts of unsupervised text data, which can be in the architectures of decode-only \cite{achiam2023gpt,touvron2023llama} or encoder-decoder \cite{xue-etal-2021-mt5}.
Since both phoneme sequence $\mathbf{h}$ and subword sequence $\mathbf{y}$ are discrete tokens, any type of LLMs, whether decode-only or encoder-decoder, can be naturally fine-tuned for P2G $p(\mathbf{y}|\mathbf{h})$.
A challenge in training and using the LLM-P2G model is that there seems to have information loss in cascading S2P and P2G. In the following, we propose two training strategies (DANP and TKM) to address this challenge.


\subsection{Data Augmentation with Noisy Phonemes (DANP)}

Usually in ASR testing, the 1-best phoneme sequence generated from the S2P model is fed into the P2G model.
So if the P2G model $p(\mathbf{y}|\mathbf{h})$ is fine-tuned using only the single annotated phoneme sequence, then there is a severe mismatch in training and testing for ASR. The input phonemes fed to P2G in ASR testing is much noiser. Therefore, a straightforward strategy to compensate for such a mismatch is to add noise to the input phonemes fed to P2G in training. There are two schemes.

\textbf{Beam search.~}We employ beam search on the S2P model to generate top-$K$ phoneme hypotheses for each utterance. 

\textbf{Random sampling.~}Alternatively, we can draw stochastic samples from the S2P model to generate more diversified hypothesized phoneme sequences. Taking the CTC-based S2P model as an example (which is the one used in our experiments), the sampling procedure is as follows.
We calculate the softmax probabilities for S2P output units including IPA symbols and blank symbols, and at frame $t=1,\cdots,T$, we draw $R$ symbols according to the softmax probabilities. Thus we obtain $R$ paths, each with length $T$.
For each path, we remove all blanks and repeated labels from the path, which is the same as in CTC to get a label sequence from a path \cite{graves2006connectionist}.
Finally, we de-duplicate the samples to obtain the hypothesized phoneme sequences.



\begin{table*}
    \centering
    \caption{ 
        Word error rate (WER) comparison for Whistle fine-tuning (FT) models and LLM-P2G models. 
        Results are shown for Polish and German, with and without language model (LM). 
        Under any column, except ``Whistle Subword FT'', the other four rows share the same acoustic model (or say S2P), called the Whistle-S2P model.        
        For Whistle models, ``w/o LM" means beam search, while ``w LM" means decoding with the WFST framework. For LLM-P2G, ``w/o LM" means beam search, while ``w LM" means using additional re-scoring with LM. NA denotes not applied.
    }
        \begin{tabular}{c|cc|cc|cc|cc}
            \toprule
            \multirow{3}{*}{\textbf{Model}} & \multicolumn{4}{c}{\textbf{Polish}} & \multicolumn{4}{c}{\textbf{German}} \\
            & \multicolumn{2}{c}{\textbf{130 h}} & \multicolumn{2}{c}{\textbf{20 h}} & \multicolumn{2}{c}{\textbf{130 h}} & \multicolumn{2}{c}{\textbf{20 h}} \\
             & \textbf{w/o LM} & \textbf{w LM} & \textbf{w/o LM} & \textbf{w LM} & \textbf{w/o LM} & \textbf{w LM} & \textbf{w/o LM} & \textbf{w LM} \\
            \midrule
            Whistle Phoneme FT    & NA & 4.30 & NA & 16.27 & NA & 15.73 & NA & 30.71\\
            Whistle Subword FT    & 5.84  & 3.82 & \textbf{17.59} & \textbf{13.84} & 14.09 & 14.01 & \textbf{27.78}  & \textbf{28.04} \\
            \midrule
            LLM-P2G        & 5.71 & 5.04 & 23.75 & 21.56 & 14.76 & 14.39 & 32.26 & 31.45 \\
            LLM-P2G + DANP   & 4.44 & 4.18 & 19.99 & 19.05 & 13.86 & 13.63 & 30.49 & 29.97 \\
            LLM-P2G + randomized TKM  & \textbf{4.01} &\textbf{3.68} & 19.19 & 17.36 & \textbf{13.44} & \textbf{13.03} & 29.20 & 28.78 \\
            \bottomrule
        \end{tabular}
    
    \vspace{-1.7em} 
\end{table*}

\subsection{Randomized top-$K$ Marginalized (TKM) Training}
The DANP strategy mainly address the mismatch in training and testing when P2G decoding is based on the 1-best phoneme sequence generated from S2P.
Ideally, in P2G decoding, it would be better to marginalize over multiple hypothesized $\mathbf{h}$ to decode according to Eq. (\ref{eq:S2P2G}). To enable matched training and testing, it would be better for the training objective of the P2G model $p(\mathbf{y}|\mathbf{h})$ to be maximizing the marginal likelihood.
For training and testing with such a latent-variable model, we are inspired by the RAG-Sequence technique used in Retrieval-Augmented Generation (RAG) \cite{lewis2020retrieval}.
Technically, it treats the retrieved document as a latent variable that is marginalized to get the marginal likelihood via a top-$K$ approximation.
In our case, the top-$K$ hypothesized phoneme sequences are generated using the S2P model, and the P2G model produces the subword sequence probability for each hypothesized phoneme sequence, which are then marginalized:
\begin{equation}
    \begin{split}
        p(\mathbf{y}|\mathbf{x}) & \approx \sum_{\mathbf{h} \in \text{top-}K(p(\mathbf{h}|\mathbf{x}))} p(\mathbf{h}|\mathbf{x})p(\mathbf{y}| \mathbf{h}) \\ 
               &= \sum_{k = 1}^K 
               p({\mathbf{h}}^{(k)}|\mathbf{x}) \prod_{i=1}^{L} p(y_i |  {\mathbf{h}}^{(k)}, y_{1:i-1})
    \end{split}
\label{eq:topk-marginal} 
\end{equation}
Specifically, ${\mathbf{h}}^{(k)}$ represents the $k$-th phoneme sequence generated by beam search using the S2P model. $p({\mathbf{h}}^{(k)}|\mathbf{x})$ can be calculated by the CTC forward-backward algorithm.
$p(y_i | {\mathbf{h}}^{(k)}, y_{1:i-1})$ can be calculated by the LLM-based P2G model in an autoregressive way.

The above training technique for latent-variable models can be generally referred to as top-$K$ marginalized (TKM) training. Furthermore, we propose a variation of TKM training where more randomness is introduced in selecting the hypothesized phoneme sequences for marginalization.
Every time the training instance $(\mathbf{x},\mathbf{y})$ is in a training minibatch, 
instead of always taking the top-$K$ hypothesized phoneme sequences (as ranked by the S2P model),
we randomly draw $n$ hypothesized phoneme sequences from top-$K$ for marginalization ($n<K$), as follows:
\begin{equation}
p(\mathbf{y}|\mathbf{x}) \approx \sum_{j = 1}^n p(\mathbf{h}^{(k_j)}|\mathbf{x})p(\mathbf{y}| \mathbf{h}^{(k_j)})   
\label{eq:random-TKM}
\end{equation}
where $k_1,\cdots,k_n$ are uniformly drawn from $1,\cdots,K$.
The training objective in Eq. (\ref{eq:random-TKM}) is referred to as randomized TKM training, to be differentiated from the standard TKM training. The advantages include 1) better generalization, as it reduces over-reliance on a specific S2P ranking; 2) more robust to noisy S2P, as it helps when real-world S2P returns imperfect results.

\subsection{Top-$K$ Marginalized (TKM) Decoding}
In decoding, we first run S2P beam search to obtain a set of top-$K$ phoneme sequences, $\mathbf{h}^{(1)},\cdots,\mathbf{h}^{(K)}$. 
Then, for each phoneme sequence $\mathbf{h}^{(k)}$, we run P2G beam search with size $S$ and, after de-duplication, we obtain a set of subword sequences $\mathbf{Y}$.
Each subword sequence in $\mathbf{Y}$ is scored by Eq. (\ref{eq:topk-marginal}).
A subword sequence $\mathbf{y} \in \mathbf{Y}$ may not have appeared in the beam decoded from every $\mathbf{h}^{(k)}, k=1,\cdots,K$.
Thus for efficient decoding, we make an approximation $p(\mathbf{y}|\mathbf{h}^{(k)}) \approx 0$, when $\mathbf{y}$ was not generated during beam search from $\mathbf{h}^{(k)}$.
This is analogous to Fast Decoding in the RAG-sequence technique \cite{lewis2020retrieval}.
Finally, based on the score Eq. (\ref{eq:topk-marginal}), the top-$S$ subword sequence in $\mathbf{Y}$ are obtained, which can be further re-scored, by combining Eq. (\ref{eq:topk-marginal}) with in-domain language model scores.


Compared to the DANP strategy, TKM decoding addresses the mismatch in training and testing when P2G decoding is based on the multiple candidate phoneme sequence generated from S2P.
Remarkably, for TKM decoding with the P2G model obtained by randomized TKM training, a similar decoding procedure can be applied, except that we run S2P beam search to obtain top-$n$ phoneme sequences, because we do not need to introduce randomness in decoding. In randomized TKM training, the P2G model is trained to infer the likely subword sequence from $n$ candidate phoneme sequences.

\section{Experiment}

\subsection{Dataset}
Experiments are conducted on the CommonVoice (CV) dataset \cite{commonvoice:2020}, version 11.0 (released September 2022). Two languages from different language families, Polish (pl) and German (de), are selected, with 130 hours of training data per language, as they both use Latin script like in the pre-trained S2P model's languages and are well-represented in the pre-trained LLM.

\subsection{Model training and testing}
Our models are trained with the CAT ASR toolkit \cite{An2020CATAC}. We use the publicly released Whistle-S model \cite{yusuyin2025whistle} as the backbone, which is a Conformer \cite{gulati2020conformer} based multilingual acoustic model, pre-trained with connectionist temporal classification (CTC) \cite{graves2006connectionist} on ten CV languages. 

We establish two baselines for each language by fine-tuning (FT) the Whistle-S backbone using weak phoneme labels\footnote{Since the IPA phonetic transcripts are obtained by the LanguageNet G2P tool, rather than from human annotations \cite{hasegawa2020grapheme}.} and subword labels, respectively.
They are denoted by ``Whistle Phoneme FT'' and ``Whistle Subword FT'' in Table 1, respectively.
The phoneme-based FT model is referred to as Whistle-S2P.
In the setting of ``w LM'', a 4-gram word-level language model is performed in both WFST-based decoding and re-scoring after beam search with $S$=4.
This crosslingual experiment setup follows Whistle \cite{yusuyin2025whistle}.

The Whistle-S2P model is used in combination with the LLM-P2G to perform the two-step ASR.
The LLM-P2G is obtained by fine-tuning the mT5-base \cite{xue-etal-2021-mt5} over the phoneme data generated by Whistle-S2P. 
mT5-base, having 583 million parameters, is an encoder-decoder Transformer \cite{vaswani2017attention}, pre-trained on the mC4 dataset (101 languages, including Polish and German).
The training details of different methods are as follows.

For applying DANP with beam search, the beam size ($K$) is set to be 1, 32 or 64 to generate noisy data ($K$-beam). For applying DANP with random sampling, we set the number of samples ($R$) to 25,000 for Polish and 500 for German. After de-duplication, the data size is augmented to about 32 times.
Then, the Whistle-S2P model is frozen and full-parameter fine-tuning is performed on mT5-base with a fixed learning rate of 3e-4 and early stopping. The resulting LLM-P2G model is denoted by ``LLM-P2G + DANP'' in Table 1. Ablation results with different hyper-parameters are shown in Table 2.

In TKM training of LLM-P2G, the hyper-parameters $K$ and $n$ are set to be 32 and 8, respectively.
The resulting LLM-P2G model is denoted by ``LLM-P2G + randomized TKM'' in Table 1. Top-8 is used in TKM decoding.
Ablation results are shown in Table 3 and 4.

\section{Result and Ablation}
\subsection{Results}
The main results are shown in Table 1.
For full training data (130 hours), the main observations are as follows: 1) LLM-P2G without DANP or TKM shows poor results, because of information loss (row 1 and 2 vs 3). 2) With DANP, for Polish, while LLM-P2G does not surpass subword fine-tuning, it reduces WER by 2.7\% compared to phoneme fine-tuning. For German, it achieves a relative WER reduction of 13.3\% and 2.7\% compared to the two baselines (rows 1 and 2 vs 4). 3) LLM-P2G with randomized TKM outperforms all other models, achieving relative WER reductions of 14.4\% and 3.6\% for Polish, and 17.1\% and 6.9\% for German (rows 1 and 2 vs 5).
The WER reductions are significant, with p-value=1e-4 (3.82 vs 3.68) and 8e-23 (14.01 vs 13.03) for Polish and German respectively, according to matched-pairs significance test {\cite{significance}}.

For low-resource (20 hours), similar trends are observed.
Randomized TKM performs the best among different LLM-P2G settings. However, only for German it achieves a reduction in relative WER of 6.2\% compared to the phoneme fine-tuning baseline (row 1 vs 5), while for Polish, it fails to achieve similar improvements. Presumably, this is because mT5-base is trained with more pre-training data for German, enabling better phoneme-to-grapheme conversion, whereas the limited Polish pre-training data restrict the performance in low-resource scenarios \cite{xue-etal-2021-mt5}. The percentages for German and Polish in mT5-base pre-training data are 3.05\% and 2.15\% respectively.
This indicates that while LLM-P2G decoding enjoys a simpler pipeline than WFST-based decoding and can leverage the pre-trained capability of LLMs, its performance depends on the amounts of pre-training data and fine-tuning data.


\subsection{Ablation study}

\begin{table}
	\centering
	\caption{
		Word error rates (WERs) for LLM-P2G with different settings of DANP.
        After de-duplication, the data size augmented by random sampling is about 32 times.
	}
    \setlength{\tabcolsep}{3.9pt} 
	\begin{tabular}{c|cc|cc}
        \toprule
         \multirow{2}{*}{\textbf{DANP strategy}} & \multicolumn{2}{c}{\textbf{Polish}} & \multicolumn{2}{c}{\textbf{German}} \\
        & \textbf{w/o LM} & \textbf{w LM} & \textbf{w/o LM} & \textbf{w LM} \\
        \midrule
	 1-beam & 5.71 & 5.04 & 14.76 & 14.67 \\
     sampling & 5.09 & 4.93 & 14.82 & 14.65 \\
     32-beam & 4.62 & 4.36 & 14.17 & 14.04 \\
     64-beam & 4.72 & 4.36 & 14.17 & 13.97 \\
     32-beam + sampling & 4.51 & 4.27 & 14.01 & 13.91 \\
	 96-beam + sampling & 4.66 & 4.26 & 13.86& 13.64 \\
     + multiple checkpoints  & \textbf{4.44} & \textbf{4.18} & \textbf{13.86} & \textbf{13.63} \\
	\bottomrule
    \end{tabular}	
	\vspace{-1.8em}
\end{table}

\begin{table}
	\centering
	\caption{
		Word error rates (WERs) for LLM-P2G with different settings of TKM training and decoding.
	}
	\begin{tabular}{c|cc|cc}
        \toprule
         \multirow{2}{*}{\textbf{TKM strategy}} & \multicolumn{2}{c}{\textbf{Polish}} & \multicolumn{2}{c}{\textbf{German}} \\
        & \textbf{w/o LM} & \textbf{w LM} & \textbf{w/o LM} & \textbf{w LM} \\
        \midrule
     top-32 & 16.55 &  16.12 & 21.69 & 21.31 \\
	 top-8 & 4.31 & 3.80 & 13.58 & 13.18 \\
     rand. 8 of top-32 & \textbf{4.01} &  \textbf{3.68} & \textbf{13.44} & \textbf{13.03} \\ 
	\bottomrule
    \end{tabular}	
	\vspace{-0.7em}
\end{table}

\textbf{DANP strategy.} Seven different data augmentation settings are used in training LLM-P2G.  
In testing, P2G decoding is based on the 1-best phoneme sequence generated from S2P, which is referred to as Best Path Decode.
The results are shown in Table 2. First, the performance improves as the amount of augmented data increases. Second, the combined use of both beam search and random sampling yields superior results, compared to using either alone. Third, to further promote diversity, we try generating noisy phonemes by 5 checkpoints saved from training, in addition to using beam search and sampling. This setting yields the best DANP result, which is also the result shown in Table 1.

\begin{table}
	\centering
	\caption{
		Comparison of word error rate (WERs) for LLM-P2G, using different training and decoding strategies. r-TKM denotes randomized TKM training. 
	}
	\begin{tabular}{c|c|c|c}
        \toprule
        \multirow{2}{*}{\textbf{Lang.}} & \multirow{2}{*}{\textbf{Train}} & \textbf{Best Path Decode} & \textbf{TKM Decode} \\
        &  & \textbf{w LM} & \textbf{w LM} \\
        \midrule
	    \multirow{2}{*}{Polish} & DANP  & 4.18  & 4.06 \\
        & r-TKM  & 3.99  & \textbf{3.68} \\
        \midrule 
        \multirow{2}{*}{German} & DANP &  13.63 & 13.59  \\
        & r-TKM & 13.42   &  \textbf{13.03} \\
	   \bottomrule
    \end{tabular}	
	\vspace{-1.8em}
\end{table}

\textbf{TKM strategy.} Different settings are examined for TKM strategy. As shown in Table 3, the setting with top-32, due to excessive noise, performs poorly in learning the correct P2G conversion. In contrast, when the number of candidates are reduced to 8, the model surpasses both the two baselines and the DANP model. Furthermore, random sampling of 8 sequences out of top-32 strategy achieves sufficient diversity while reducing noise, and obtains the best performance, which is the result shown in Table 1 (row 5).

\textbf{TKM training and decoding.} We train LLM-P2G using DANP or randommized TKM (abbreviated as r-TKM), and compare the performance of each model with best path decoding (top-1) or TKM decoding (top-$K$). As shown in Table 4, compared to DANP training, r-TKM training achieves relative WER reductions of 4.5\% and 9.3\% for Polish, and 1.5\% and 4.1\% for German, using best path decoding and TKM decoding, respectively (row 1 vs 2 and 3 vs 4). Similarly, compared to best path decoding, TKM decoding reduces relative WER by 2.8\% and 7.7\% for Polish, and 0.2\% and 2.9\% for German, under DANP and r-TKM training respectively (column 3 vs 4). These results demonstrate that the TKM method improves both model training and decoding performance.


\section{Conclusion}

In this paper, we propose LLM-P2G for phoneme-based ASR, which belongs to a two-step ASR architecture, consisting of speech-to-phoneme and LLM-based phoneme-to-grapheme. Moreover, by incorporating data augmentation with noisy phonemes (DANP) and randomized top-$K$ marginalized (TKM) training and decoding, we effectively mitigate performance degradation caused by potential information loss in cascading S2P and P2G models. Our experimental results demonstrate that LLM-P2G not only outperforms WFST-based ASR systems for crosslingual ASR but also simplifies the decoding pipeline. The new method is efficient and flexible in leveraging acoustic and linguistic knowledge from large pre-trained models, offering a promising direction for future research.


\ifinterspeechfinal
\else
\fi

\bibliographystyle{IEEEtran}


\end{document}